\documentstyle[aps,prb,preprint]{revtex}
\tightenlines
\title{Thermopower and thermal conductivity of superconducting perovskite $MgCNi_3$}
\author{S.Y. Li, W.Q. Mo, M. Yu, W.H. Zheng, C.H. Wang, Y.M. Xiong, R. Fan, H.S. Yang, B.M. Wu,  L.Z. Cao, and X.H. Chen$^{\ast}$} 
\address{Structural Research Laboratory and Department of Physics, 
University of Science and Technology of China,  Hefei, Anhui 230026, 
P. R. China}
\date{\today}
\begin{document}

\maketitle

\begin{abstract}

The thermopower and thermal conductivity of superconducting perovskite $MgCNi_3$ ($T_c \approx$ 8 K) have been studied. The thermopower is negative from room temperature to 10 K. Combining with the negative Hall coefficient reported previously, the negative thermopower definetly indicates that the carrier in $MgCNi_3$ is electron-type. The nonlinear temperature dependence of thermopower below 150 K is explained by the electron-phonon interaction renormalization effects. The thermal conductivity is of the order for intermetallics, larger than that of borocarbides and smaller than $MgB_2$. In the normal state, the electronic contribution to the total thermal conductivity is slightly larger than the lattice contribution. The transverse magnetoresistance of $MgCNi_3$ is also measured. It is found that the classical Kohler's rule is valid above 50 K. An electronic crossover occures at $T^* \sim 50 K$, resulting in the abnormal behavior of resistivity, thermopower, and magnetoresistance below 50 K. 

\end{abstract}   
 
\vskip 10 pt


\section*{introduction}

Recently the discovery of two new intermetallic superconductors $MgB_2$\cite{Nagamatsu} and $MgCNi_3$\cite{He} has attracted great attention. $MgB_2$ has relatively high superconducting transition temperature ($T_c$ = 39 K) and the highly promising potential application. High frequency phonons induced by the light element, B, are believed to be essential in yielding the high $T_c$. Despite its low $T_c$ (about 8 K), $MgCNi_3$ is of interest because it is the first compound having the perovskite structure without any oxygen, also bacause of the high proportion of Ni in it. 

$MgCNi_3$ forms a three-dimentional perovskite structure. Comparison to a familiar oxide perovskite such as $CaTiO_3$, for example, indicates the structural equivalencies between Ca and Mg, Ti and C, and O and Ni. Instead of Ti atom, C atom is located in the body-centered position surrounded by a $Ni_6$-octahedra cage. Such a large amount of nickel in $MgCNi_3$ make it an analog of the three-dimentional layered nickel borocarbides, typified by $LuNi_2B_2C$ with a $T_c$ near 16 K.\cite{Hilscher} One might expect nickel to give rise to possible magnetism in this compound, but no long range magnetic ordering transition was observed from 2 K to 295 K by neutron-diffraction measurements.\cite{Huang} Band calculations indicate predominantly Ni 3d character at the Fermi level of $MgCNi_3$,\cite{Dugdale,Shim,Singh} similar to $LuNi_2B_2C$\cite{Mattheiss,Pickett} and $YNi_2B_2C$.\cite{Lee} Ni-derived 3d electrons are considered to be superconducting electrons in $MgCNi_3$, so the possibility of the existence of a localized moment on Ni atoms is not reasonable since it would lead to strong pair breaking if superconductivity is due to $s$-wave pairing. He {\it et al}.\cite{He} has determined the electron-phonon coupling constant $\lambda_{ph} \sim 0.77$ by specific heat measurement, which is in the range of conventional phonon. Tunneling measurements by Mao {\it et al}.\cite{Mao} detected a zero bias conductance peak below $T_c$ suggesting an uncoventional character of the superconducting pairing state. However, the recent $^{13}C$ NMR data\cite{Singer} below $T_c$ are consistent with isotropic $s$-wave superconductivity.  

There are some doping investigations on $MgCNi_3$,\cite{Hayward,Ren} but only few transport properties have been reported. Our previous Hall effect measurement\cite{Li} indicates that the carrier in $MgCNi_3$ is electron-type. The value of $R_H$ at 100 K agrees well with that calculated by Singh {\it et al}.\cite{Singh} However, their calculated thermopower is $p$-type except for that at very low $T$. They found that $S$ is very small (less than 1 $\mu$V/K) below 150 K, then rises more rapidly reaching 5 $\mu$V/K at 300 K. They contributed this unusual $T$-dependence of $S$ to the competition between the hole and electron pockets of comparable size in the Fermi surface. 

Thermal conductivity $\kappa$ is one of those transport coefficients which exhibits non-zero values in both the normal and the superconducting state. The temperature dependence of $\kappa$ allows one to distinguish between the most important interactions in a superconductor. In particular, the interaction of electrons with phonons is recorded in the magnitude of $\kappa(T)$. Moreover, scattering of these particles by static imperfections like impurities, defects or grain boundaries is reflected.

In this paper, we report the first measurement of thermopower and thermal conductivity for superconducting perovskite $MgCNi_3$. The thermopower is negative from room temperature to 10 K. Combining with our previous Hall effect measurement, it definetly indicates that the carrier in $MgCNi_3$ is electron-type. The origin of the nonlinear temperature dependence of $S(T)$ below 150 K is discussed. The thermal conductivity is of the order for intermetallics, larger than that of borocarbides and smaller than $MgB_2$. In the normal state, the electronic contribution to the total thermal conductivity is slightly larger than the lattice contribution. The transverse magnetoresistance of $MgCNi_3$ is also measured. It is found that the classical Kohler's rule is violated below 50 K. The abnormal behavior of resistivity, thermopower, and magnetoresistance below 50 K is attributed to an electronic crossover occuring at $T^* \sim 50 K$. 

\section*{experiment}

Polycrystalline specimen of $MgCNi_3$ used for present measurements is the same one reported previously.\cite{Li} Detailed preparation has been reported in Ref. 15. The thermopower was measured by a dc method in the temperature range 10 and 300 K. The temperature gradient ($\Delta T$) in the sample was measured using the two pairs of Rh-Fe thermocouples. The sample was mounted on the top of two well separated copper blocks with silver paint. During the measurement, the temperature gradient $\Delta T$ of two separated copper blocks was kept at 1 K. To eliminate the effects of the reference leads, the absolute thermopower of copper was subtracted from the measured thermoelectric voltage. The thermopower result displayed in this paper is the averaged value over five data measured at the same temperature. 

The method of longitudinal steady-state thermal flow was used in measuring the thermal conductivity of the sample from 7 to 300 K. There is a resistive heater at one end of the sample, while the other end makes good heat exchange with a heat source whose temperature can be set freely. A Rh-Fe thermometer is used on the heat source, and the temperature gradient of the sample is measured by a NiCr-Constantan thermocouple. We use PID arithmetic to control the temperature by computer. The temperature stability is better than $\pm$2 mK, which is available to the precise requisition of thermal measurement.

Magnetoresistance was measured by a standard four-probe method. The magnetic field was applied by superconducting magnet system (Oxford Instruments) and the temperature was measured by a Cerox thermometer.

\section*{results and discussion}

Fig. 1 shows the X-ray diffraction (XRD) pattern and temperature dependence of resistivity under zero magnetic field for the $MgCNi_3$ sample. XRD pattern indicates that the sample is nearly single phase. Rietveld refinement\cite{Mo} has been performed based on the space group Pm-3m. It gave the cubic cell parameter $a$ = 3.81111 \AA\ and actual C occupancy about 0.97, being consistent with previous report.\cite{He} The resistive superconducting transition is very sharp with the midpoint of the resistive transition $T_c$ = 8.0 K. The temperature dependence of resistivity, $\rho(T)$, is almost identical to that reported by He {\it et al}.\cite{He} A similar $\rho(T)$ behavior has been observed in $(Ba,K)BiO_3$ thin film\cite{Moon} and single crystal.\cite{Affronte} Above 70 K, the normal state $\rho(T)$ behavior follows Bloch-Gr\"{u}neisen theory consistently with electron-phonon scattering and the fitting curve is shown as solid line in Fig. 1.\cite{Li} 

\subsection*{A. Thermopower}

The absolute thermopower $S(T)$ of $MgCNi_3$ from room temperature (RT) to 10 K is shown in Fig. 2. The upper inset of Fig. 2 shows the temperature dependence of Hall coefficient reported by us.\cite{Li} $S(T)$ is negative for the whole temperature range, which is in contrast to that predicted by Singh {\it et al}.\cite{Singh} The negative thermopower, combining with the negative Hall coefficient as seen in the upper inset of Fig. 2, definetly indicates that the charge carrier in this compound is electron-type. Room-temperature thermopower $S(RT)$ and $[dS/dT]_{RT}$ are -9.2 $\mu$V/K and -13.1 nV/K$^2$, respectively. The magnitude of $S(RT)$ is somewhat larger than the typical value associated with free electron/conventional metals, i.e., -1.28 $\mu$V/K for lead and 1.94 $\mu$V/K for gold, but it is approximately the same as for palladium\cite{Barnard} [$S(RT)$ = -10 $\mu$V/K] and $Y(Lu)Ni_2B_2C$ single crystals.\cite{Rathnayaka}     

The thermopower of conventional nonmagnetic metals consists of two contributions, a diffusion contribution and a phonon-drag contribution resulting from the transfer of phonon momentum to the electron gas. The diffusion contribution is proportional to temperature, while the phonon-drag contribution falls at low temperature as the phonons freeze out, and  at high temperatures as the excess phonon momentrum gets limited by phonon-phonon scattering. This usually results in a phonon-drag peak in conventional metals with $T^3$ dependence below 0.1$\Theta_D$ and falls as $T^{-1}$ above $\approx$0.3$\Theta_D$. From Fig. 2, no obvious phonon-drag peak is present from RT down to 10 K in $MgCNi_3$. 

The thermopower of $MgCNi_3$ is approximately linear in $T$ near room temperature within the measurement accuracy. The extrapolation of the $S(T)$ data, assuming a linear $T$ dependence of $S$, does not pass through $S$ = 0 at $T$ = 0 and gives large intercept. The data in the temperature region between 150 K and RT was fitted to a straight line, i.e., $S(T) = a + bT$, with $a$ = -5.2 $\mu$V/K and $b$ = -13.1 nV/$K^2$. The intercept value is slightly larger than that reported for single crystals of $YNi_2B_2C$ (-4.6 $\mu$V/K) and $LuNi_2B_2C$ (-4.3 $\mu$V/K).\cite{Rathnayaka} The thermopower of $MgCNi_3$ shows the change in slope at the "knee" at about 150 K. Obviously, in addition to the diffusion thermopower, there is(are) additional contribution(s) to the thermalpower yielding the observed temperature dependence of $S$. Such a nonlinear temperature dependence of thermopower $S(T)$ is very similar to that of $Y(Lu)Ni_2B_2C$ single crystals.\cite{Rathnayaka} Rathnayaka {\it et al}.\cite{Rathnayaka} found that $S-bT$, representing contributions to thermopower other than the diffusion thermopower, is negative and almost constant between 100 and 300 K, and varies approximately as $T^{-1}$ below 100 K for both $YNi_2B_2C$ and $LuNi_2B_2C$ single crystals. They suggest that the observed slope change of $S(T)$ in $Y(Lu)Ni_2B_2C$ single crystals could be due to the phonon-drag effect as in high-$T_c$ cuprate supercondcutors.\cite{Trodahl} We plot $(S-bT)$ vs $T$ for $MgCNi_3$ in the lower inset of Fig. 2. It is found that $S-bT$ is almost constant above 150 K, approximately -5.2 $\mu$V/K. However the data of $S-bT$ can not be fitted as $T^{-1}$ between 10 and 150 K. This result indicates that the additional contribution(s) to the thermopower in $MgCNi_3$ may not be attributed to phonon-drag as in $Y(Lu)Ni_2B_2C$ single crystal\cite{Rathnayaka} and high-$T_c$ cuprate supercondcutors.\cite{Trodahl} It is believed that the particular thermopower behavior of high-$T_c$ cuprates and borocarbides relates to their layered nature.\cite{Rathnayaka,Trodahl} $MgCNi_3$ is a three-dimentional, not layered compound, hence a different contribution to the thermopower is reasonable.

A low temperature "knee" in $S(T)$ would also be expected from electron-phonon interaction renormalization effects.\cite{Kaiser1} Electron-phonon renormalization would lead to an enhanced thermopower that is given by
\begin{equation}
   S = S_{b}[1 + \lambda(T)]
\end{equation}
where $\lambda(T)$ is the electron-phonon mass enhancement parameter and $S_b$ is the bare thermopower (without renormalization effects). In this expression certain other corrections have been ignored which are relatively small and can be ignored as a first approximation.\cite{Kaiser2} Equation (1) can be rewritten as
\begin{equation}
   {S\over T} = {S_{b}\over T}[1 + \lambda(T)]
\end{equation}
where $\lambda(T)$, the electron-phonon mass enhancement parameter, is maximum at $T$ = 0 K and becomes smaller as $T$ is raised, becoming almost negligible near RT and higher temperature in comparison with 1. A plot of $S/T$ vs $T$ should then give a measure of $\lambda(T)$, and $[S/T]_{T\rightarrow{0}}/[S/T]_{RT}$ should approximat $1 + \lambda(0)$. 

Fig. 3 shows the plot of $S/T$ vs $T$ from 10 K to RT for $MgCNi_3$. The magnitude of $S/T$ increases smoothly as temperature decreased above about 50 K, but shows a rapid increase at 50 K and a negative peak at 35 K. This nagative peak reflects the abnormal $T$-dependence of thermopower blow 50 K, as seen from Fig. 2. The electrical resistivity data of Fig. 1 also shows a change of curvature in the same temperature range, and satisfies $\rho \sim T^{n}$ with $n \sim 1.7$ below 50 K. Singer {\it et al}.\cite{Singer} found that NMR Knight shift $^{13}K$ saturates below about 50 K. They suggest that an electronic crossover takes place near $T^{*} \sim 50 K$ prior to the superconducting transition at $T_c$ = 7.0 K. This electronic crossover may be responsible for the abnormal behavior of thermopower and resistivity blow 50 K. 

Due to the presence of the negtive peak at 35 K, it is difficult to determine the ratio $[S/T]_{T\rightarrow{0}}/[S/T]_{RT}$ precisely. However, to get some qualitative feeling as to the importance of these renormalization effects in $S(T)$ for $MgCNi_3$, the $S/T$ values at 10 K and the peak (35 K) are taken to be $[S/T]_{T\rightarrow{0}}$ as an approximation, respectively. Using these values of $[S/T]_{T\rightarrow{0}}$ and value of $[S/T]_{RT}$ from Fig. 3, estimated values of $\lambda(0)$ are 1.4 and 1.7, respectively. Such a range of $\lambda(0)$ values of $MgCNi_3$ are close to that of some strong-coupling $A$-15 superconductors, e.g., $Nb_3Sn$ ($\lambda \approx 1.8$) and $Nb_3Al$ ($\lambda \approx 1.5$). The value of $\lambda(0)$ for $Y(Lu)Ni_2B_2C$ single crystals obtained from a similar analysis of $S/T$ data\cite{Rathnayaka} is unrealistically high in comparison with the values of $\lambda(0)$ for conventional superconductors including strong-coupling ones. Consequently, electron-phonon renormalization effects do not explain the $S(T)$ data for $Y(Lu)Ni_2B_2C$ single crystals.  

He {\it et al}.\cite{He} has determined the electron-phonon coupling constant $\lambda_{ph} \sim 0.77$ by specific heat measurement for $MgCNi_3$. This value of $\lambda$ seems much smaller than that of $\lambda(0)$ estimated from the $S(T)$ data. Therefore there are maybe some mechanisms other than electron-phonon coupling to induce mass enhancement. NMR measurements\cite{Singer} reveal that $MgCNi_3$ has a moderate strong ferromagnetic spin fluctuation and the spin fluctuations are nearly a factor 3 enhanced with decreasing temperature. In the presence of spin fluctuations, Eq. (2) is modified to
\begin{equation}
   {S\over T} = {S_{b}\over T}[1 + \lambda(T) + \lambda_{sf}]
\end{equation}
where $\lambda_{sf}$ is the mass-enhancement parameter due to spin fluctuations. For our present study, the sum of $\lambda$ and $\lambda_{sf}$ is expected to explain the apparently large $\lambda(0)$ obtained from the [$S/T$] vs $T$ plot.

\subsection*{B. Thermal Conductivity}

Fig. 4 shows the temperature dependence of the thermal conductivity $\kappa$ of $MgCNi_3$. $\kappa(T)$ keeps nearly constant above 210 K. Below that temperature, the positive slope of $\kappa(T)$ indicates the limitation of the heat conductivity by crystal defects as in pure normal metals $\kappa$ exhibits a maximum at lower temperatures and then decreases with rising temperature. The magnitude of $\kappa$ is of the order for intermetallics, larger than that of borocarbides\cite{Sera} and smaller than $MgB_2$.\cite{Bauer,Schneider} As seen from the inset of Fig. 4, $\kappa$ shows  a small but clear decrease at $T_c$ = 8 K, which agrees well with that determinded from the resistivity measurement. 

Generally, the total thermal conductivity of metals consists of an electronic contribution $\kappa_e$ and a lattice contribution $\kappa_l$:
\begin{equation}
   \kappa = \kappa_{e} + \kappa_{l}
\end{equation}
In order to separate both contributions from the total measured effect, the Wiedemann-Franz law\cite{Berman} is applied, assumed to be valid, at least, in simple metals. This model relates the electrical resistivity $\rho$ with the electronic contribution to the thermal conductivity $\kappa_e$, and  $\kappa_e$ can be expressed as 
\begin{equation}
   \kappa_{e}(T) = {L_0T \over {\rho(T)}}
\end{equation}
where $L_0 = 2.44 \times 10^{-8} W{\Omega}K^{-2}$ is the Lorenz number.

Using Eq. (4) and taking into account the appropriate values of the normal-state resistivity of $MgCNi_3$ allows to split $\kappa$ into $\kappa_e$ and $\kappa_l$ (dashed lines, Fig. 4). It is found that the electronic contribution to the total thermal conductivity is slightly larger than the lattice contribution in the normal state of $MgCNi_3$.

According to Matthiessen's rule both $\kappa_e$ and $\kappa_l$ are limited owing to various scattering processes, which can be expressed in terms of a thermal resistivity $W$. In the case of non-magnetic materials, the following temperature dependence of the electronic contribution to the total measured quantity is assumed to be valid:\cite{Klemens}
\begin{equation}
   1/\kappa_e(T) \equiv W_e(T) = W_{e,0}(T) + W_{e,ph}(T) = {\alpha \over T} + \beta T^2
\end{equation}
where the subscripts (e,0) and (e,ph) refer to interactions of the conduction electrons with static imperfection and thermally excited phonons, respectively; $\alpha$ and $\beta$ are material constants.

Eq. (6) allows to detemine $W_{e,0}$ and $W_{e,ph}$.  The electronic thermal resistivity $W_e$ is shown in Fig. 5 from 10 to 50 K. The solid line is the fitting curve according to Eq. (6). and the dashed lines represent $W_{e,0}$ and $W_{e,ph}$, respectively. Thus, the deduced parameters are $\alpha = 51.36$ $mK^2W^{-1}$ and $\beta = 1.2 \times 10^{-4}$ $mW^{-1}K^{-1}$. Obviously from Fig. 5, the scattering of electrons with static imperfections of the crystal becomes dominant as the temperature approaches $T_c$. 

Above discussions on thermal conductivity of $MgCNi_3$ did not take account of the effect of spin fluctuations, which have been greatly enhanced with decreasing temperature.\cite{Singer} In fact, the scattering of electrons with spin fluctuations will decrease the thermal conductivity, especially at low temperature. However, the exact temperature dependence of the electronic thermal resistivity caused by the interactions of the conduction electrons with spin fluctuations in $MgCNi_3$ is not clear for us so far. To clarify it, further experimental and theoretical works should be done.

\subsection*{C. Magnetoresistance}

To get further insight into the charge transport, the magnetoresistance (MR) measurement is a useful tool since it is more sensitive to the change in the charge carrier scattering rate 1/$\tau$, effective mass $m^*$, and the geometry of the Fermi surface. In conventional metals, the electrical conductivity can be described in terms of the Boltzmann equation.\cite{Ziman} In the presence of a magnetic field $H$, the change in the distribution function $g({\bf v})$ is described by

\begin{equation}
   g({\bf v}) = [1 + (H\tau){e\over c}{\bf v} \times {\bf \hat{H}}\cdot{{\partial v}\over {\hbar}\partial {\bf k}}\cdot {\partial \over {\partial v}}]^{-1}[-\tau{\bf {eE}}\cdot {\bf v}{{\partial f^0}\over {\partial \epsilon}}]
\end{equation}

The magnitude of the magnetic field contributes to Eq. (7) in a product of $H$ and $\tau$. Since 1/$\tau$ is generally proportional to the zero-field resistivity $\rho_0$, the MR $\Delta\rho/\rho_0$ depends only on $H/\rho_0$. This results in a scaling law referred to as Kohler's rule which holds in many conventional metals:\cite{Pippard}

\begin{equation}
   {\Delta\rho \over {\rho_0}} = f(H\tau) = F({H\over {\rho_0}})
\end{equation} 

in the low-field limit, the MR quadratically depends on $H$, and is therefore scaled as $\Delta\rho(T) = const \times (H/\rho_0)^2$.

Fig. 6 shows the normal-state transverse ($H\bot I$) magnetoresistance of $MgCNi_3$ as a function of magnetic field at various temperature. The magnetoresistance is defined as $MR = \Delta\rho/\rho_0 = (\rho(H)-\rho_0)/\rho_0$. The inset of Fig. 6 is the magnetic field dependence of the resistance at 2.3 K. Above $H_{c_2}(2.3 K)$ = 12.5 T, the resistance incresases with the magnetic field, showing a positive magnetoresistance. The change in the resistance at 12 T and 15 K is about 0.9\% which is comparable with normal metals.\cite{Pippard} A large in-plane MR ($H\|c$) ($\approx 7.3\%$) has been observed in $LuNi_2B_2C$ single crystal for $H$ = 4.5 T at 20 K, and a similar amount of MR for $YNi_2B_2C$ single crystal.\cite{Rathnayaka} The value of MR for $LuNi_2B_2C$ polycrystalline sample\cite{Narozhnyi} is about 3.5 times higher than that of single crystal. Obviously the MR of $MgCNi_3$ is much smaller than that of borocarbides. As seen in Fig. 6, the normal-state transverse MR of $MgCNi_3$ is always positive and monotonically decreases with increasing temperature. The magnetic field dependence is essentially $H^2$ up to 12 T for all temperatures. The data in Fig. 6 are replotted as $\Delta\rho/\rho$ vs $(H/\rho)^2$, Kohler's plot, in Fig. 7. Above 50 K the data fall onto a single straight line, which implies that the MR is essentially scaled by $H/\rho$, i.e., that it follows the classical Kohler's rule. The data below 50 K have a different slope, which means the deviation from Kohler's rule. As mentioned above, an electronic crossover at $T^{*} \sim 50 K$ has been revealed by NMR measurements\cite{Singer} and affects the low temperature behavior of resistivity and thermopower. It is reasonable to contribute the deviation of MR from Kohler's rule blow 50 K to this electronic crossover.

\section*{conclusion}

We have investigated the thermopower and thermal conductivity of superconducting perovskite $MgCNi_3$. Combining with the negative Hall coefficient reported previously, the negative thermopower definetly indicates  the electron-type carrier in $MgCNi_3$. The nonlinear temperature dependence of thermopower  below 150 K is explained by the electron-phonon interaction renormalization effects. The thermal conductivity is of the order for intermetallics, larger than that of borocarbides and smaller than $MgB_2$. A small but clear decrease of thermal conductivity is observed at $T_c$ = 8 K. In the normal state, the electronic contribution to the total thermal conductivity is slightly larger than the lattice contribution. The transverse magnetoresistance of $MgCNi_3$ is also measured. It is found that the classical Kohler's rule is valid above 50 K. Abnormal behavior of resistivity, thermopower, and magnetoresistance below 50 K are observed, and may be related to an electronic crossover occuring at $T^* \sim 50 K$.

\section*{acknowledgment}

This work is supported by the Natural Science Foundation of China and by the Ministry of Science and Technology of China (NKBRSF-G19990646). 

\vskip 10 pt

\noindent $\ast$ To whom all correspondence should be addressed. Email: chenxh@ustc.edu.cn

\newpage
\noindent
{\bf FIGURE CAPTIONS} \\

\noindent
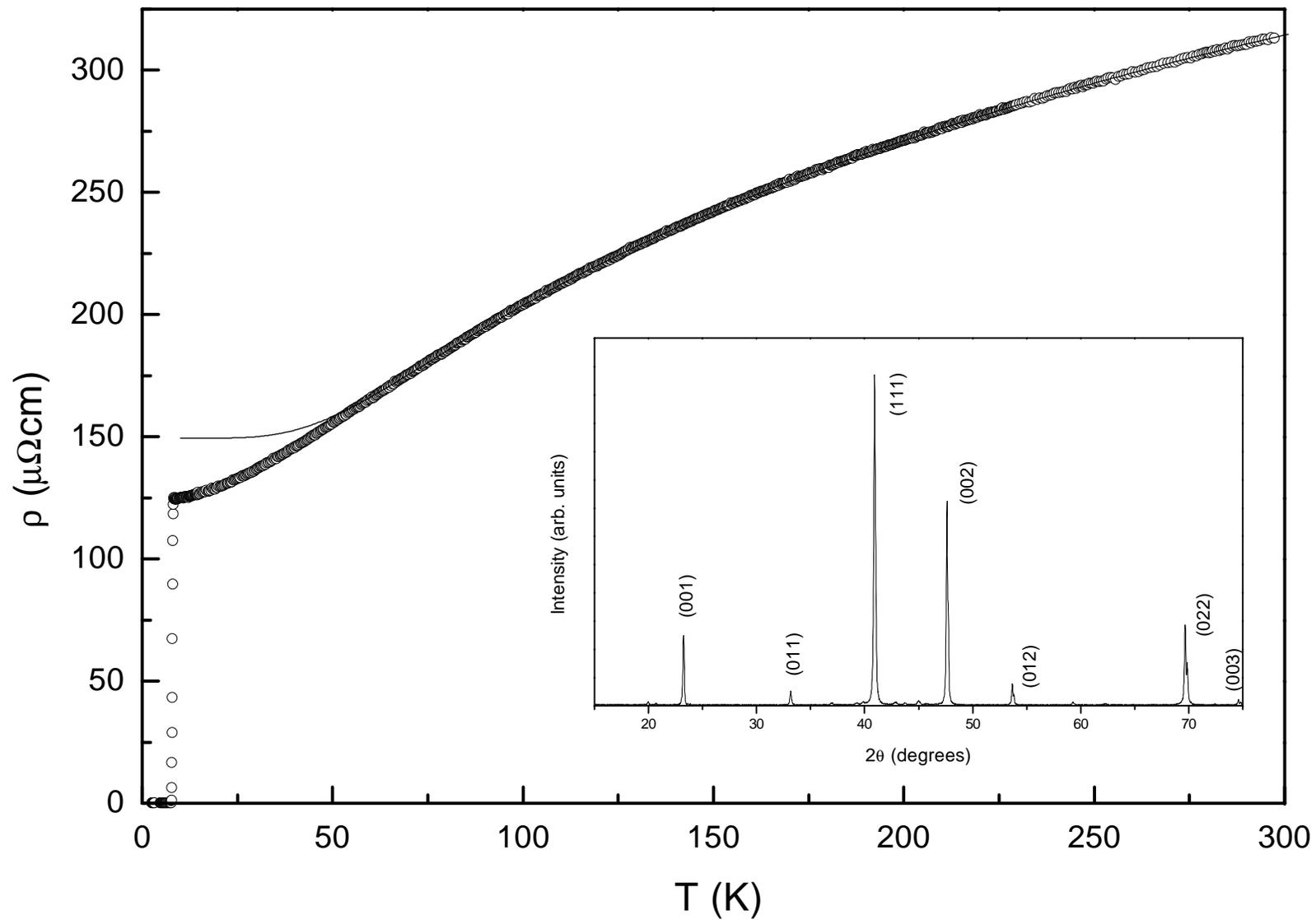
FIG 1: 
  
The temperature dependence of resistivity under zero magnetic field for $MgCNi_3$ sample. Inset: the XRD pattern.\\

\noindent
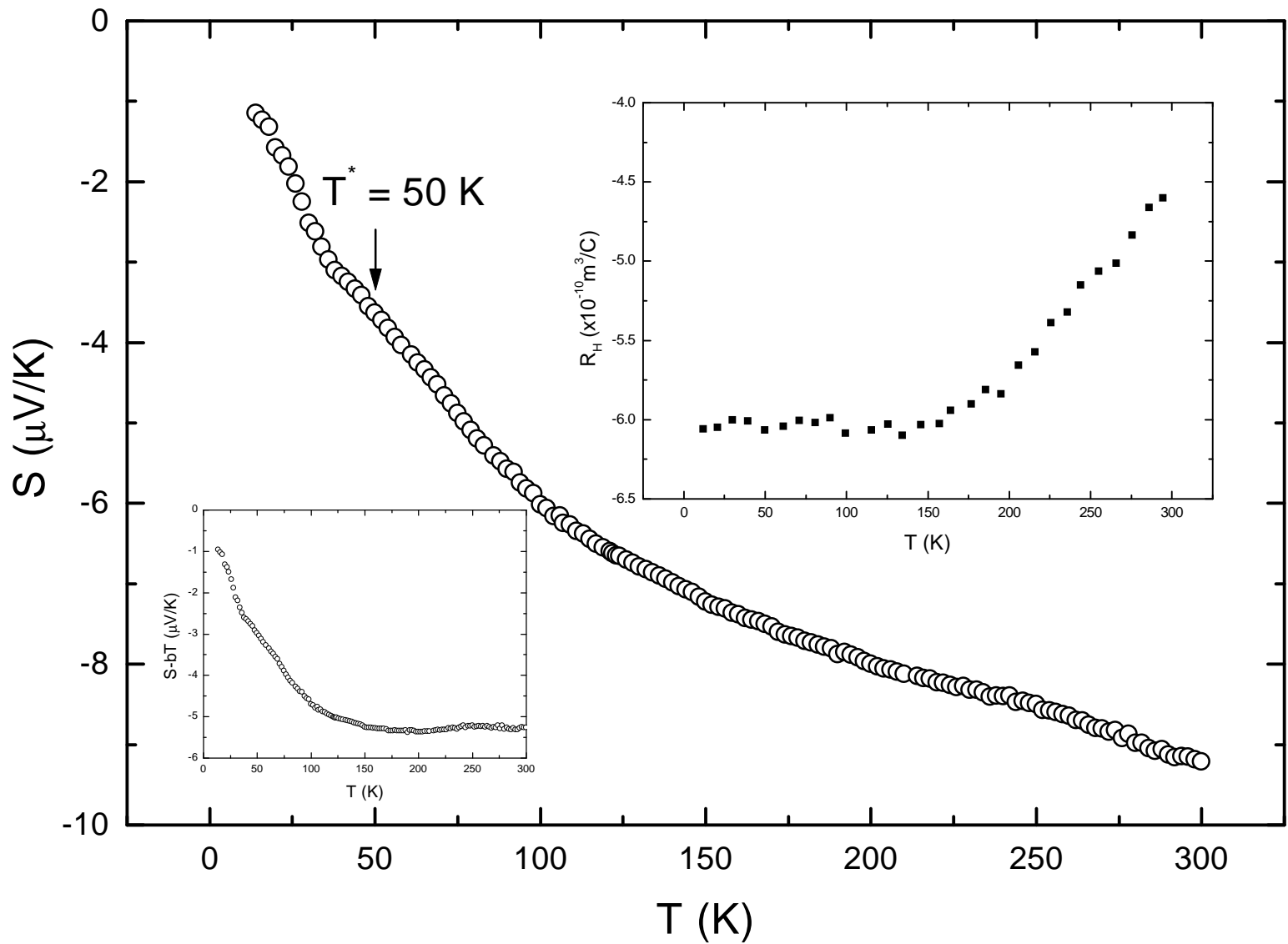
FIG 2:

The temperature dependence of thermopower for $MgCNi_3$ sample. Upper inset: the temperature dependence of Hall coefficient adoped from Ref. 15. Lower insert: $S-bT$ as a function of temperature. \\

\noindent
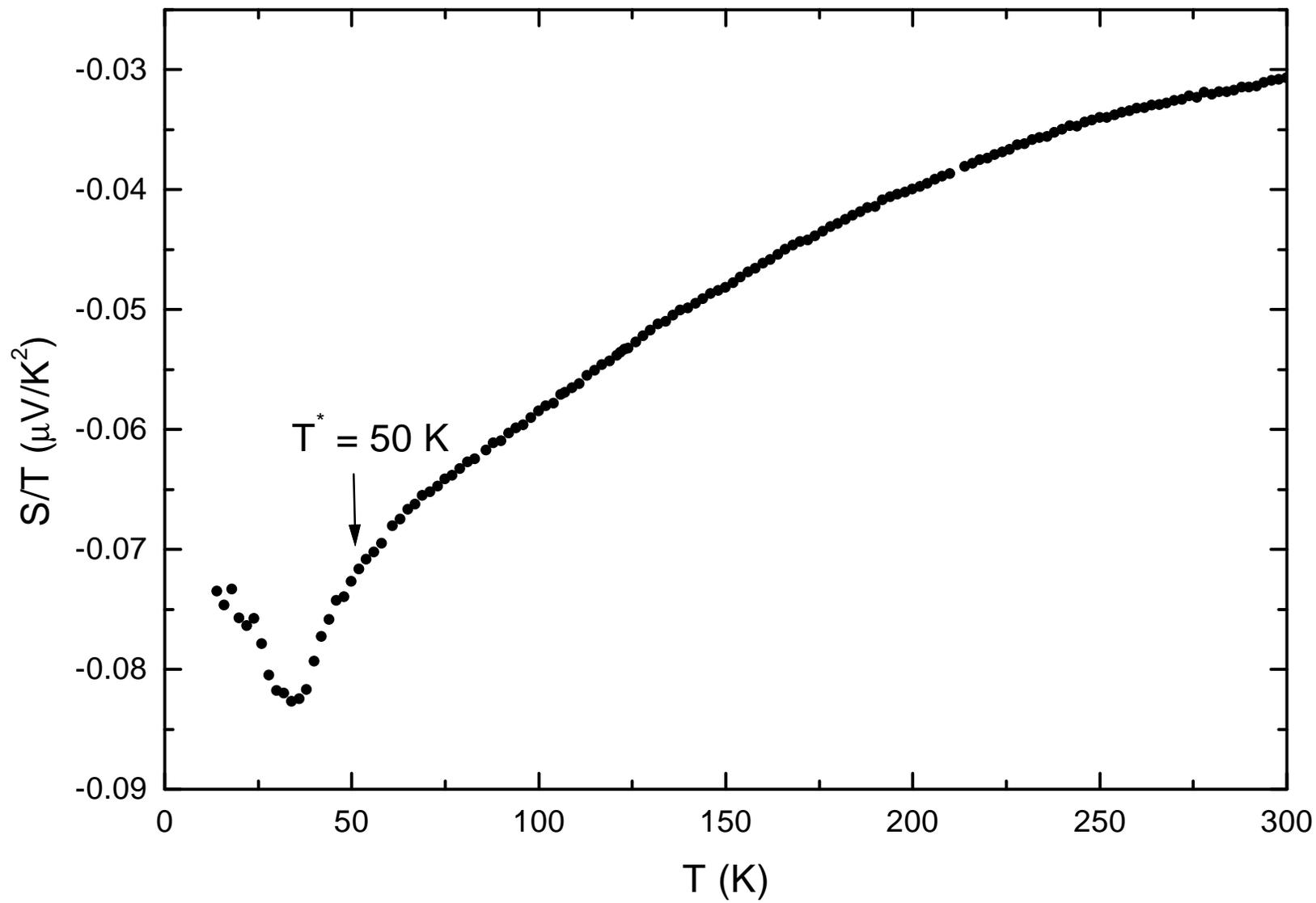
FIG 3:

$S/T$ as a function of temperature for $MgCNi_3$.\\

\noindent
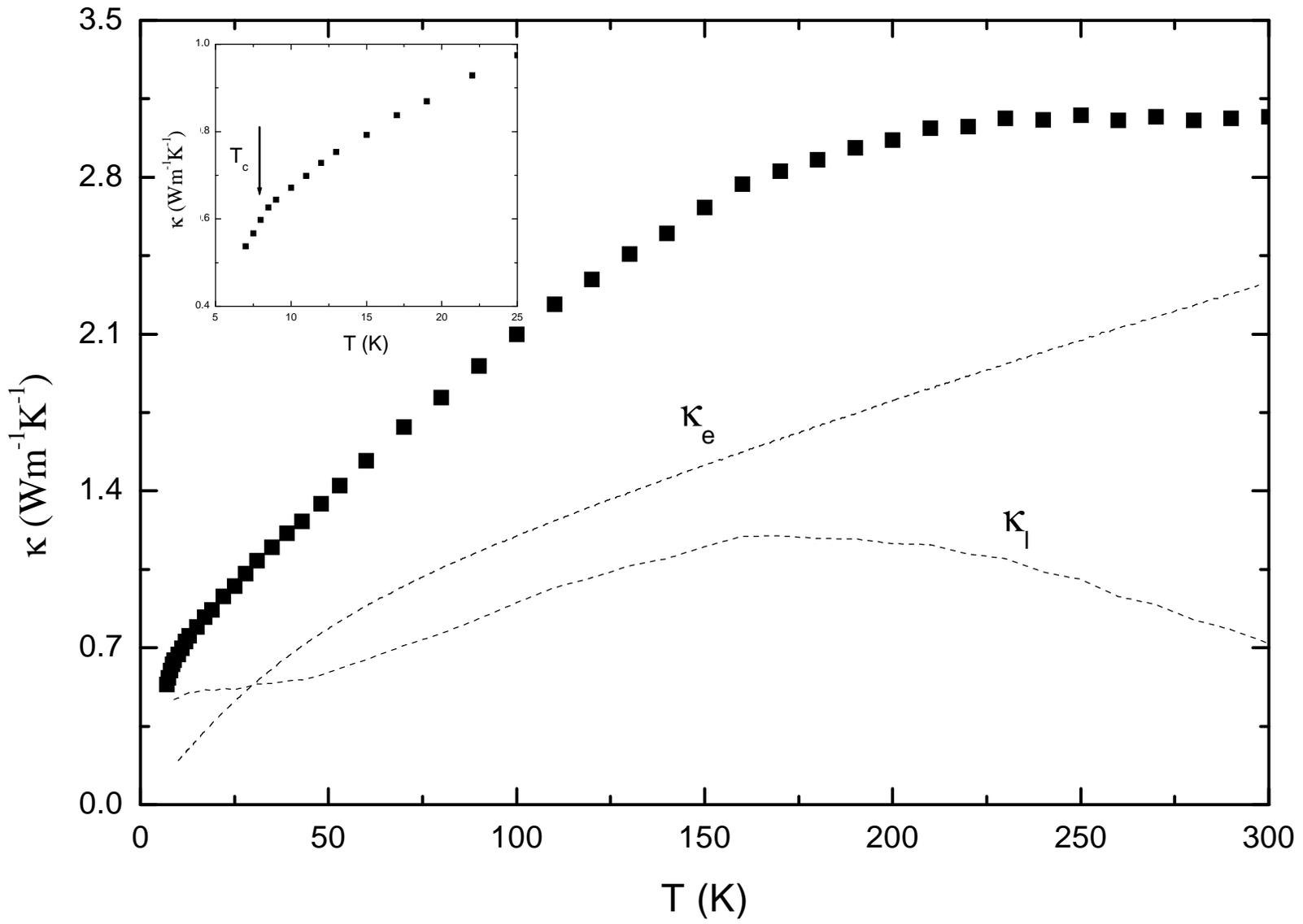
FIG 4:

The temperature dependence of the thermal conductivity of $MgCNi_3$. The dash lines represent the electronic contribution $\kappa_e$ and lattice contribution $\kappa_l$, respectively. The inset shows the change of slope at $T_c$ = 8 K.\\

\noindent
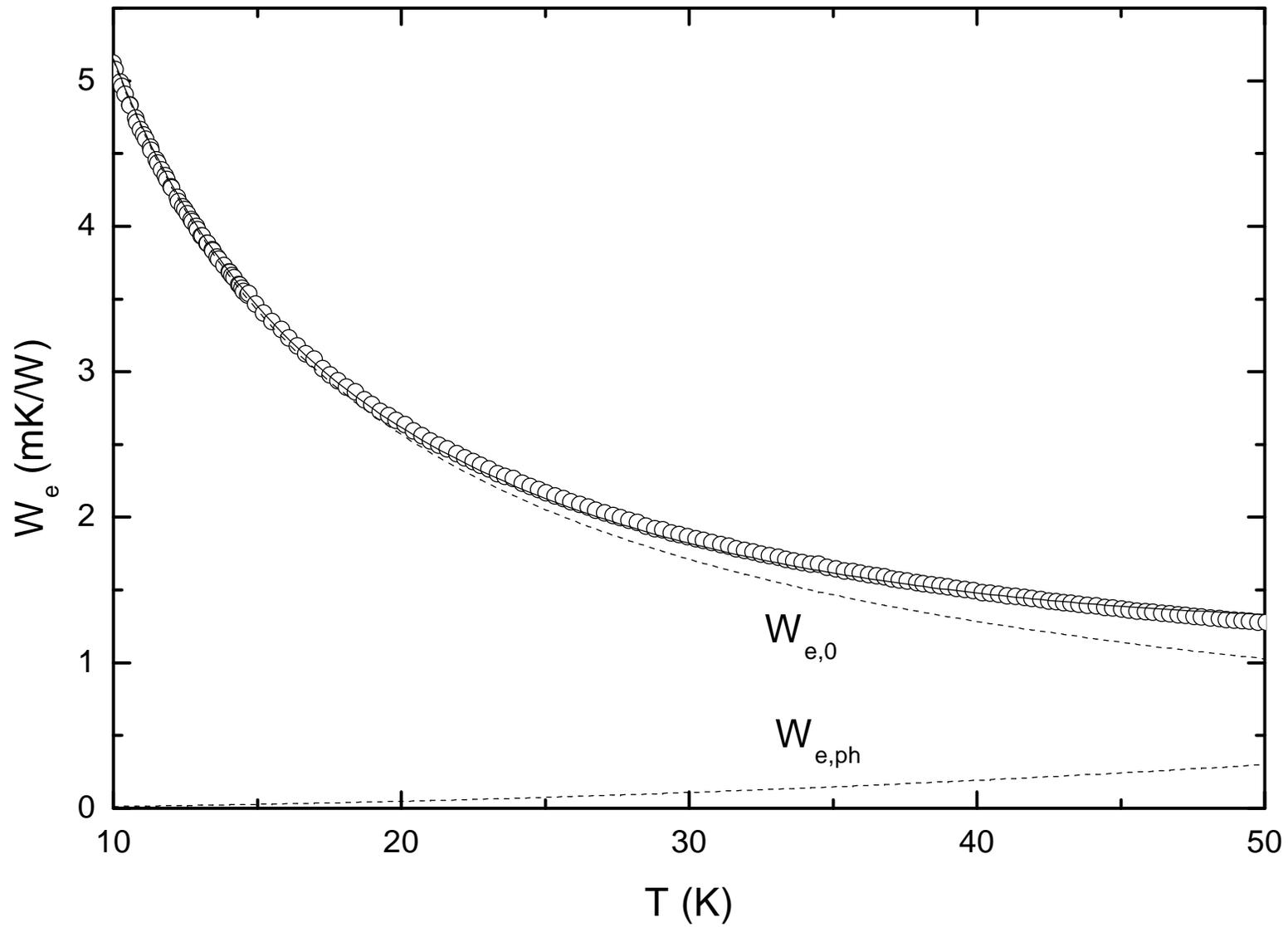
FIG 5:

The electronic thermal resistivity of $MgCNi_3$. The solid line is the fitting curve according to Eq. (6).\\

\noindent
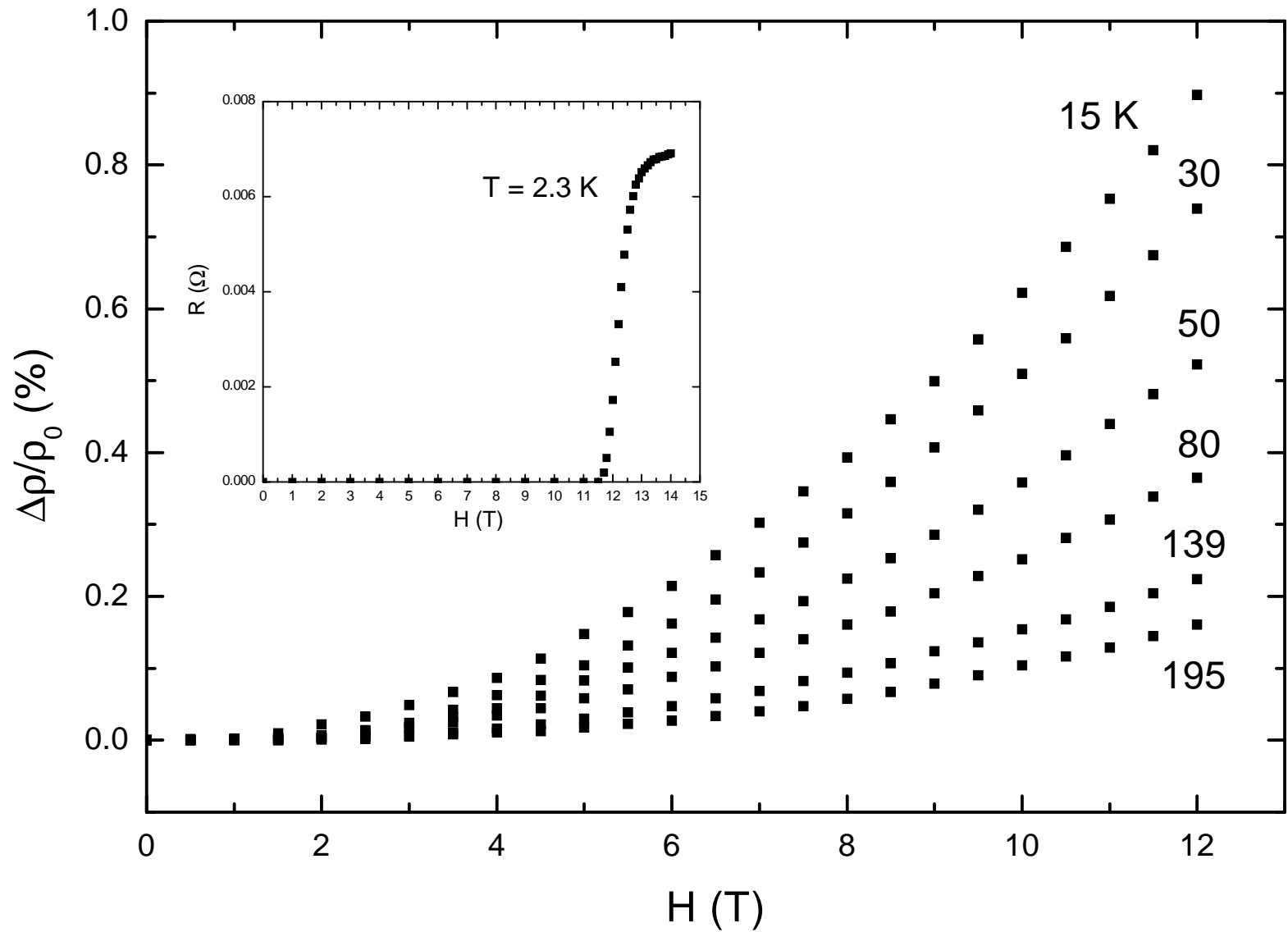
FIG 6:

The normal-state transverse ($H\bot I$) magnetoresistance of $MgCNi_3$ as a function of magnetic field at various temperature. Inset: the magnetic field dependence of the resistance at 2.3 K.\\

\noindent
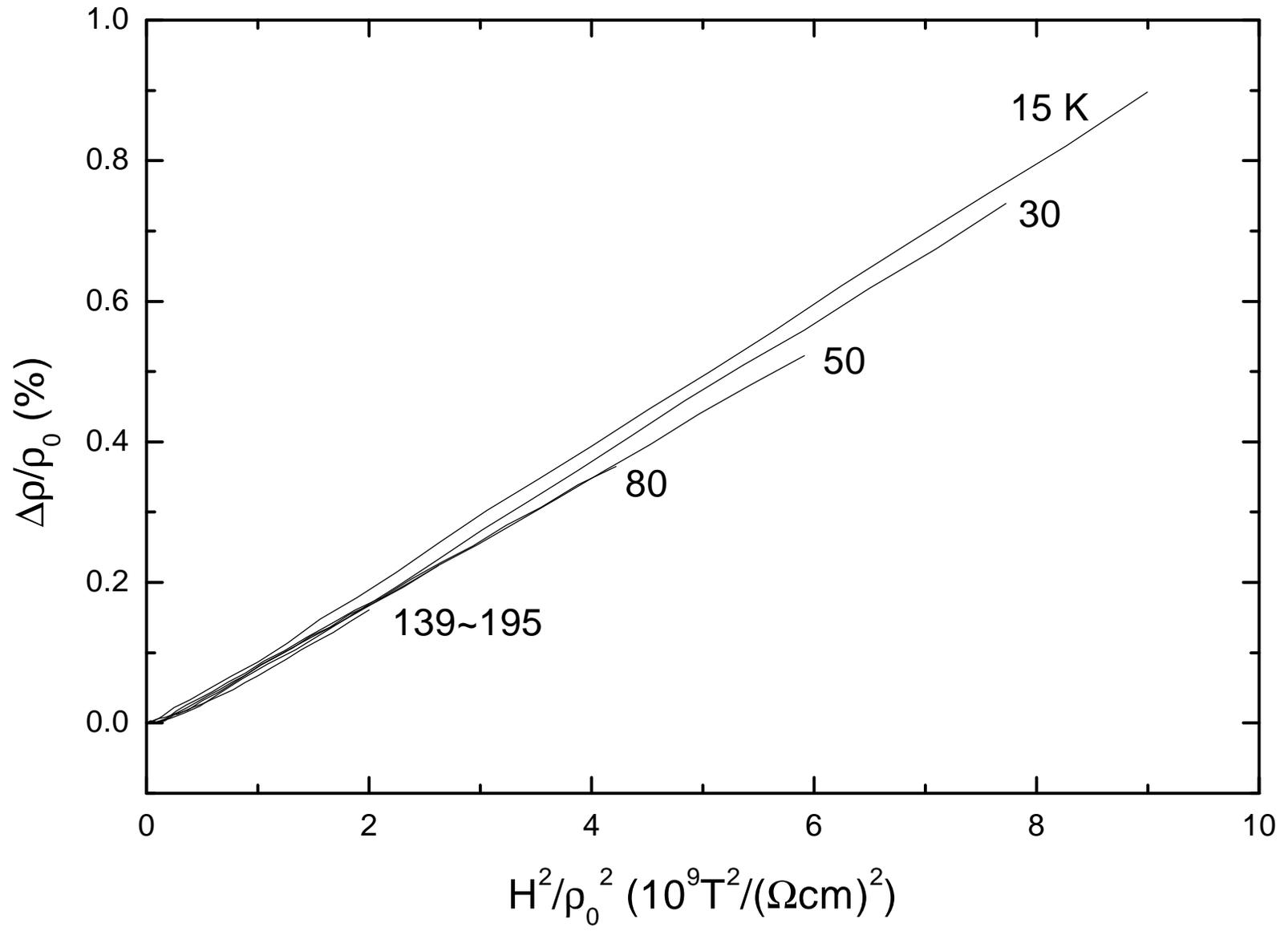
FIG 7:

Kohler's plot for $MgCNi_3$ at selected temperatures. Above 50 K the presence of a universal line implies that Kohler's rule holds.\\

\end{document}